\documentclass[11pt]{article}
\usepackage{times}
\usepackage{geometry}
\usepackage[utf8]{inputenc}
\usepackage{enumitem,amssymb}
\usepackage{graphicx}
\usepackage{fancyhdr}
\usepackage{aas_macros}
\usepackage{mdframed} 
\usepackage{breakurl}
\usepackage[breaklinks]{hyperref}
\usepackage[superscript,biblabel]{cite}

\mdfdefinestyle{theoremstyle}{
innertopmargin=\topskip,}
\mdtheorem[style=theoremstyle]{lrptextbox}{}

\begin{document}

\begin{center} {\LARGE Small and Moderate Aperture Telescopes for Research and Education}

Aaron C.~Boley$^{1,*}$, Terry Bridges$^2$, Paul Hickson$^1$, Harvey Richer$^1$, Brett Gladman$^1$, Jeremy Heyl$^1$, J.~J.~Kavelaars$^3$, Ingrid Stairs$^1$

1) The University of British Columbia, 2) Okanagan College, 3) NRC Herzberg

$^*$aaron.boley@ubc.ca

\end{center}

\begin{center} Executive Summary \end{center}

In this white paper (WP), we highlight several examples of small and moderate aperture telescopes that are being used for education and/or research.  We further discuss potential costs for establishing new, small observatories, as well as joining existing international consortia.  The WP includes a brief overview of select observing sites, with a discussion on how small telescopes at exceptional observing locations can be competitive, under certain circumstances, with larger and more expensive facilities located at poorer sites. Furthermore, small and moderate telescope facilities enable on-demand observing and/or unique observing opportunities that may not be possible at large (and often oversubscribed) telescopes.  While not meant to be all-encompassing, we discuss research programs associated with CanCON, Dragonfly Telephoto Array, ESA's Flyeye, Evryscope, Next-Generation Transit Survey, The Plaskett and the DAO 1.2 metre, the University of British Columbia Southern Observatory, the WIYN and SMARTS 0.9 metres, and the Zwicky Transient Facility. Science programs advanced by these facilities address, for example, exoplanets, binaries, variability, Solar System minor bodies, low-surface brightness structures, spectroscopy, microlensing, and large sky surveys.  We further discuss commercial telescope pay-for-service options through iTelescope.  In addition to research, these facilities enable many different types of educational experiences for wide range of people, from high school students to undergraduates to graduate students to postdocs. Potential uses include direct observing by upper-level astronomy undergraduates and graduate students and incorporation of observing and citizen science into large service courses.  Maintenance itself necessitates physically working with a research telescope and offers unique student training opportunities -- astronomy undergraduates at UBC participated in the USO's  installation at Cerro Tololo Inter-American Observatory, for example. 
Canada should remain committed to partnering with large, international observatories such as CFHT, Gemini, and TMT, but it should also negotiate international agreements and commit funding to expand the use of small and moderate research observatories at domestic and international sites through coordination with the NRC, the Tri-Council, and the Canadian Foundation for Innovation. Both capital and operational costs (with site rental costs allowed) need to be included in support possibilities.  CASCA should establish and maintain a small to moderate telescope expression of interest database that would help to facilitate Canadian institutions in organizing consortia, particularly for smaller institutions.  The astronomical community should work with the NRC to make existing facilities more accessible to the astronomical community for research. This could involve, for example, automating the Plaskett and/or providing travel funds for supporting classical observing modes. Finally, a small to moderate aperture facility in the Arctic would be a world-class observatory and should be advanced over the next decade.

\newpage


\begin{center}\bf Scope of the White Paper  \end{center}

In this white paper (WP), we present a summary of select observing facilities that use small to moderate aperture telescopes for research and education.  
For this discussion, we do not intend to define a strict aperture size that would classify a telescope as being either small or moderate, as there are many factors to consider.  Rather, we are concerned with telescopes that could be operated by a single university or a small consortium of institutions, with the exception being an Arctic telescope, which is itself only briefly discussed. The largest aperture telescope noted in this WP is the Plaskett 1.8 metre, and is included, in part, as a potential facility that could be modernized through modest investment.  The aperture size should not be taken as a general indicator of the complexity of an observing facility, either, as a modern, compact one-metre telescope can be operationally simpler to use and maintain than, for example, an array of small telescopes.  

In choosing telescopes to discuss, we have tried to show a variety of approaches, sites, and science objectives, but we do not mean to be all-encompassing.  An omission of a telescope or a site location in this WP should not be taken as a negative outlook for that facility in any way.    We further realize that there are many different ways that small or medium telescopes could be used for education, including direct use, indirect use, hands-on work (maintenance and instrument design), remote observing, site visits, etc., and we only seek to highlight some of the possibilities. 

During the discussion, we also mention several companies, which is intended to demonstrate some of the commercial options for observatory components or services. This is not intended to be an endorsement of those companies.  

This WP is organized as follows.  We first highlight examples of different types of small and moderate aperture telescopes, as well as the diversity of science programs they are conducting.   We then give a brief summary of several select site locations that are used to host telescopes for research and/or education and have the capacity for establishing new facilities or have potential partnership opportunities. We then emphasize that in certain circumstances, smaller telescopes at premier observing locations can be competitive with larger telescopes at poorer sites.  Several estimates for cost considerations are discussed next, followed by a brief overview of options for incorporating small research telescopes into astronomy curricula.  We conclude by providing recommendations for the Long Range Plan.  

On a final note, one of the purposes of this WP is to emphasize the utility in small and moderate observatories and to argue that they can be supported, in a variety of ways, along with Canada's involvement in large, world-class facilities.

\begin{center}\bf A short list of small to moderate telescope observatories and some of their science programs \end{center}

CanCON (Canadian Collaborative Occultation Network) is a network of small telescopes (a mix of portable 30 cm Sky-Watcher telescopes, and fixed-pier telescopes with diameters ranging from 28 to 46 cm) equipped with QHY CMOS cameras, located at seven sites in BC's Okanagan valley and one site in Victoria. Each site is run by high school teachers and their students and/or RASC members. CanCON is a Canadian extension of the US RECON project\cite{tnorecon_net}, and was launched in September 2018. Its main goals are to determine the sizes of objects in the outer Solar System through stellar occultations, to engage youth in astronomical research, and to enable collaborations between high school students and teachers, amateur astronomers, and professional astronomers. 

The Dragonfly Telephoto Array\cite{abraham_dokkum_2014} consists of forty-eight 400 mm f/2.8 commercially available telephoto lenses that function together to observe low surface brightness objects.  The array is equivalent to a one-metre aperture telescope with f/0.4. The principal science goal is to ``improve our understanding of dark matter through study of the low surface brightness universe''\cite{dragonfly_web}, and is able to investigate, for example, the halos of spiral galaxies, low luminosity galaxy groups, ultra-diffuse galaxies\cite{dokkum_etal_2016}, and faint field dwarf stars.  The array can also be used to study exoplanet transits during bright time.  The telescope, hosted by the New Mexico Skies\cite{nmskies} telescope facility, is supported by the NSF, Dunlap Institute, NSERC, and NRC, with the Co-Is being located at the Dunlap Institute, Yale University, and Harvard University\cite{dragonfly_web}.

The European Space Agency's (ESA) Flyeye\cite{flyeye_esa} is a planned one-metre telescope that will split the incoming light into 16 different optical paths and detectors designed to work together to provide a 45 sq.~deg field of view (FOV).  When operational, the telescope will observe 50\% of the hemisphere three times per night using 30 to 40 s exposures, with principal objectives being near Earth object (NEO) detection for planetary defence and space debris monitoring for enhanced space situational awareness. The limiting magnitude is expected to be approximately 21.5 mag, with the hope of detecting asteroids down to 40 metre in diameter with sufficient time to react to a local impact risk\cite{jehn_flyeye,torralba_etal_2019}.

Evryscope is a small observatory that consists of twenty-four 61 mm telescopes that can observe an 8000 sq.~deg. FOV simultaneously in each two-minute exposure\cite{evryscope_design}.  Two of these telescope arrays are planned, with Evryscope south already in operation at the Cerro Tololo Inter-American Observatory (CTIO) in Chile.  The science goals of the telescope focus on ``capturing rare, short-timescale events across the sky''\cite{law_everyscope_2015}.  The system is specifically designed to be comparable to large surveys, but focusing on targets brighter than 16th magnitude, with applications to stellar variability, transiting exoplanets, eclipsing binaries, accreting compact objects, and microlensing events\cite{law_everyscope_2015}.  The system is expected to provide early coverage of other bright phenomena, such as supernovae and optical components of gamma-ray bursts. Evryscope is funded by the NSF and operated by The University of North Carolina, Chapel Hill.

In comparison with academic facilities, iTelescope\cite{itelescope} is a private company that provides observing under a pay-for-service model.  The company owns facilities in Australia, California, Chile, New Mexico, and Spain, with telescope apertures ranging from about 10 to 60 cm. Instrument and observing options are intended to cater to many types of telescope users, including amateur astronomy, education, and research.  We will discuss iTelescope further below for a comparison between observatory ownership and the company's fee model.  

Next-Generation Transit Survey\cite{wheatley_etal_2018} uses an array of twelve 200 mm f/2.8 Astro Systeme Austria telescopes, with a combined FOV of 96 sq.~deg. The plate scale is $4.97''$ per pixel, and the photometric precision is reported to routinely meet 150 ppm for $\rm V<13$ mag.  The primary science objective of the survey is to discover Neptune-sized planets around Sun-like stars and super-Earths around M dwarfs\cite{wheatley_etal_2018}, with demonstrated results\cite{west_etal_2019}.  The facility is located at Cerro Paranal Observatory in Chile, and is operated by a consortium of institutes with support from the UK Science and Technology Facilities Council and in-kind contributions from Paranal Observatory. Capital costs ``were funded by the University of Warwick, the University of Leicester, Queen's University Belfast, the University of Geneva, the Deutsches Zentrum f\"ur Luft- und Raumfahrt e.V.~(DLR; under the `Gro{\ss}investition GI-NGTS'), the University of Cambridge and the UK  [STFC]''\cite{wheatley_etal_2018}.

The Plaskett (1.8 metre) and the 1.2 metre are located at the Dominion Astrophysical Observatory (DAO) in Saanich, BC, offering imaging and spectroscopy.  The telescopes, which are oversubscribed, support approximately 12 programs per year among 20-35 PIs and Co-Is (D.~Bohlender, private communication). Research programs include, for example, minor planet observations, WD eclipses, SNe spectroscopy, binary stars, magnetic fields in faint peculiar Ap and Bp stars, and abundance studies.  The 1.2 metre can be operated robotically, while the 1.8 metre still requires an operator.  The telescopes also serve the local community for astronomy outreach events.  Refereed papers using data from these telescopes have ranged from 12 to 21 per year between 2007 and 2016.    

The UBC Southern Observatory (USO) is located at CTIO in Chile, operated under contract with Association of Universities for Research in Astronomy (AURA).  Positioned on a tower at the mountain summit, it has some of the best seeing at the site (typically less than $1''$).  The current telescope is a 35 cm Cassegrain, but is being updated to a Planewave 50 cm Corrected Dall-Kirkham (CDK) thanks to an NSERC RTI grant.  The facility is robotic, and can be operated remotely.  The USO will support on-demand observing, student training, Geostationary Earth Orbit (GEO)  debris research, exoplanet system monitoring, and minor body observing.

The WIYN\cite{WIYNp9}  and the SMARTS\cite{SMARTS}  0.9 metres are twin Cassegrain telescopes located on Kitt Peak and Cerro Tololo, respectively.  The University of Wisconsin, Indiana University, Yale University, and NOAO operate the WIYN telescope, along with additional partners, while SMARTS is run by an open consortium.  Both telescopes have only classical observing mode, are used for direct imaging (although the detectors are different), and conduct research and HQP training.  The SMARTS consortium further operates a 1.0, 1.3, and 1.5 metre,  all at CTIO, and typically have led to over 10 papers per year.

The Zwicky Transient Facility (ZTF) is a 47 sq.~deg FOV camera mounted on a 48-in Schmidt telescope at Palomar Observatory, and is included, in part, as an example of a facility that has undergone multiple successful upgrades.  The observatory seeks to ``produce a photometric variability catalog with nearly 300 observations each year, ideal for studies of variable stars, binaries\cite{burdge_etal_2019}, AGN, and asteroids [and comets\cite{bodewits}]''\cite{ztf}. The telescope scans nearly 4000 sq.~deg per night at a depth of about 20.5 mag, and is supported by ``the National Science Foundation  including Caltech, IPAC, the Weizmann Institute for Science, the Oskar Klein Center at Stockholm University, the University of Maryland, the University of Washington, Deutsches Elektronen-Synchrotron and Humboldt University, Los Alamos National Laboratories, the TANGO Consortium of Taiwan, the University of Wisconsin at Milwaukee, and Lawrence Berkeley National Laboratories''\cite{ztf}.

\begin{center}\bf Comments on Select Observing Sites \end{center}

Site quality can vary significantly, even among major observatories\cite{hellemeier_etal_2019}.  The choice of a site will depend on the planned science and education objectives, the convenience of the facility, and operational costs. As before, the following are intended to highlight different locations at which current facilities could be modified, new observatories could be built, or existing consortia could be joined, but is far from being all-encompassing.   

For commercial options, iTelescope offers telescope hosting at its locations, as does New Mexico Skies\cite{nmskies}.  Most of the sites have lower seeing quality compared with, for example, Cerro Tololo, and may have a lower fraction of photometric nights. Based on the information provided on their websites, the mountain conditions for many of their locations would be comparable to those found on Mount Kobau\cite{kobau}.  Science cases that do not require excellent seeing (e.g, Dragonfly Telephoto Array, which is located on New Mexico Skies infrastructure), could be operated from these sites. Many of the locations offer 24/7 support. 

The Plaskett and 1.2 metre are situated on Observatory Hill at DAO.  The atmospheric seeing and lower number of photometric nights are limitations, but the location remains productive (see above) and the existing infrastructure could be used for additional projects, provided support staff demands can be met.  Moreover, Canadian institutions, DAO, and the NRC could coordinate to make the Plaskett and the 1.2 metre more widely accessible to the Canadian astronomical community for research and HQP training, which could include automation of the Plaskett and/or travel support to the facility.  

Cerro Tololo and Kitt Peak, operated by AURA, have opportunities for adding observatories or joining existing consortia as an international partner.  In some cases, existing telescopes need new owners to assume operational control and responsibility. Canadian consortia could use such situations to expand observing programs to international locations, including those in the southern hemisphere.  Cerro Tololo in particular has excellent seeing conditions, which can have a significant effect on detection limits (discussed below).

An Arctic telescope was noted in the LRP2010, but establishing such a facility remains elusive. Several studies have demonstrated that PEARL on Ellesmere Island could be an ideal site for circumpolar observations in the high Canadian Arctic\cite{law_arctic_2013,hickson_arctic}, with long-term photometric monitoring possible during winter and orbital debris research during brighter nights.  Even a small to moderate size telescope could make significant research contributions, and international partnerships with Antarctic teams could be coordinated.  The high costs of operating in the Arctic would likely require NRC involvement along with a university consortium.  Should orbital debris research be included, the DND could also be involved. 

The point of the above series of comments is to emphasize that there are a variety of location options (domestic and international) for advancing science that can be conducted with small to moderate-sized telescopes, as well as for use with students and HQP. 

\newpage

\begin{center}\bf Sensitivity Estimates for a Small Telescope \end{center}

As already mentioned, the site constraints are set, in part, by the science and/or education objectives of a given telescope.  For example, should a telescope primarily be used for education, then reliably good weather and  easy accessibility (whether physically, virtually, or both) may be the principal considerations.  Wide-field surveys that utilize large detector pixels may require dark skies, but not necessarily excellent seeing. 

Here, we point out that even a small telescope at an observatory with excellent seeing conditions has the potential to outperform moderately-large telescopes for detection studies.

Let the detector counts for a given source in the V band with magnitude $V$ be given as
\begin{equation}
S = 1000~ BW Q_E ~Q_S ~A (1-\eta) ~10^{-0.4 V} ~T ,
\end{equation}
where $BW$ is the bandwidth for the filter in angstroms, $Q_E$ is the quantum efficiency for the detector,  $Q_S$ is the local sky transmission, $A$ is the telescope collecting area in sq.~cm, and $\eta$ is the fractional area blockage due to the secondary and/or other support structures. The integration time is $T$. The factor of 1000 comes from the definition of a $V=0$ mag source.   These counts will compete with the sky background, which can be written as
\begin{equation}
BG = 1000~ BW ~Q_E ~Q_S ~A (1-\eta) ~10^{-0.4 \mu_V} ~T  ~\pi \rm (FWHM)^2. 
\end{equation}
The quantities have a similar meaning, except we now use the sky surface brightness $\mu_V$.  We further use aperture photometry for our detection, with an aperture that is twice the seeing disc ($2\times\rm FWHM$ in arcseconds).  Using these definitions, the signal to noise ratio (SNR) is estimated by 
\begin{equation}
SNR = \frac{S}{\sqrt{S+BG+\sigma_{RN}^2}}
\end{equation}
for readnoise $\sigma_{RN}$.

\begin{figure}[h]
\begin{center}
\includegraphics[width=4in]{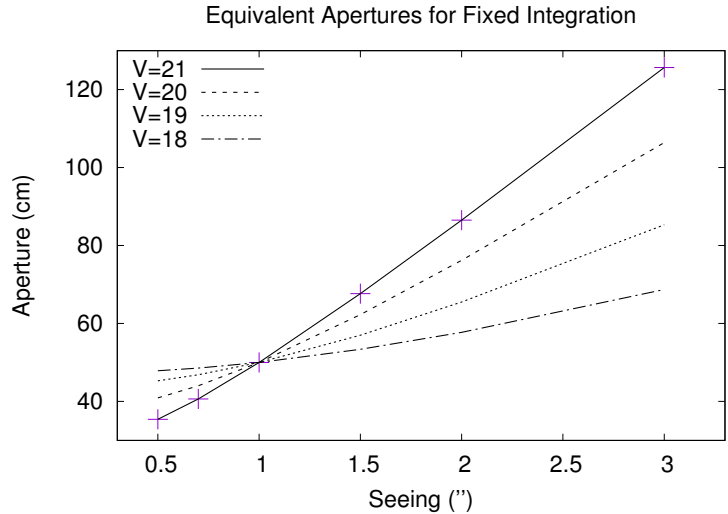}
\caption{Aperture sizes needed to reach the same SNR, as a function of astronomical seeing, for a given $V$ magnitude point source at fixed integration $T$, assuming read noise and dark current are negligible.  We have further assumed that all telescopes use the same detectors and filters, that $\eta$ is fixed, and the sky conditions are the same with $\mu_V=21.5$ mag per sq.~arcsec.  Thus, only the seeing varies for each curve. The crosses emphasize apertures at $0.5''$, $0.7''$, $1.5''$, $2''$ and $3''$ for the faint $V=21$ mag source.  For bright point sources, the effects of seeing on detection become marginalized; but, for faint objects, a small telescope under good seeing conditions can be competitive with a larger telescope under poor seeing conditions.  } 
\end{center}
\end{figure}

Now, consider the following situation: A telescope has an aperture of 50 cm and a blockage $\eta=0.157$.  
The detector is characterized by $Q_E=0.61$ (such as for an FLI Proline) using the V band with $BW =1000 \rm \AA $ and read noise $\sigma_{RN}=5$ counts.  
Further suppose that the sky brightness $\mu_V=21.5~\rm mag$ per sq.~arcsec, $Q_S=0.88$, and the seeing $\rm FWHM=1''$, such as Cerro Tololo under moderate seeing conditions.    
A source at $V=21$ mag  can be detected with an $\rm SNR=10$ after a $T\approx 90$ s integration.  All else being equal, observing with seeing at $\rm FWHM\approx 2''$ would require a $T\approx 260$ s integration.

We can further compare the faint point source detection capabilities of small telescopes under good seeing conditions against larger telescopes under poorer seeing conditions. Assuming all telescopes have the same detector and filter and that the sky background and transmission are the same, the aperture needed to detect a source at $V$ for a fixed integration time can be written as
\begin{equation}
D_1 = D_0 \left(\frac{1 + 10^{-0.4(\mu_V-V)}\pi ({\rm FWHM}_1)^2}{1 + 10^{-0.4(\mu_V-V)}\pi ({\rm FWHM}_0)^2}\right)^{1/2},\label{Dratio}
\end{equation}
for a reference aperture $D_0$ and seeing $\rm FWHM_0$. A caveat is that we have assumed dark current and read noise are negligible, which will not necessarily be the case.  Figure 1 shows the results using $D_0=50$ cm  and ${\rm FWHM}_0=1''$, $V=21$ mag, and $\mu_V=21.5$ mag per sq.~arcsec, highlighting that an 86 cm aperture under $2''$ seeing can detect a faint point source as well as a 41 cm aperture under $0.7''$ conditions.  Said differently, a 50 cm telescope at CTIO will meet or outperform the DAO 1.2 metre under the typical seeing conditions for each site, at least for faint point source detection.  

We stress that the above calculations are only one example of many considerations for choosing a site and/or a telescope.  The calculations further assume that the detector has a small enough pixel scale to take advantage of better seeing.  Observatories built to detect extended, low surface brightness objects may not  benefit significantly from being at a location with low atmospheric turbulence (e.g., Dragonfly). The sky conditions (except seeing) were also kept fixed for these comparisons.  As the point source becomes much brighter than the background, the effect is reduced (Fig.~1) and eventually eliminated (eq.~\ref{Dratio}).

\begin{center}\bf Funding Considerations  \end{center}

The cost of a small or moderate aperture research and teaching observatory is well within the grasp of a large university or a consortium of smaller institutions. Moreover, the cost of a robotic telescope at a premier observing site is comparable to or less than many physics research labs.   Here, we consider the capital and operational costs of small and moderate-sized telescopes, using several existing facilities as an example, as well as the ongoing upgrade of the UBC Southern Observatory.

Capital costs: First, there are multiple telescope and dome providers.  We focus on two of each.  Planewave offers several observatory telescopes, including a 17", 20", and 24" on direct-drive mounts in either equatorial or alt-az configurations.  They also offer a 70 cm and a one-metre telescope, both of which are integrated into an alt-az mount.   Their products include software that can be controlled through TCP/IP, as well as through an interface.  The listed price for a 20" CDK and mount is \$50,000 USD, but can be closer to \$60,000 USD when including recommended options.  There are many configuration changes for the larger offerings that create a jump in price, with the 70 cm listed at \$210,000 USD and the one-metre at \$650,000 USD.  

Another company is Astro Systeme Austria (ASA).  ASA offers a 400 mm Ritchey-Cr\'etien and mount for about 40,000 EUR including multiple options. As with Planewave, control software is included.  They further advertise an alt-az 80 cm ``turn-key'' system with an integrated mount and enclosure.  The 80 cm alt-az configuration on its own is quoted around 210,000 EUR, and their one-metre option and mount is just over 500,000 EUR, comparable to Planewave's.  ASA further has 1.2, 1.5, and 2.0 metre options, also integrated with mounts.    

AstroHaven and AstroShell offer clamshell enclosures, with several size options, that can be controlled remotely/robotically.  Such designs remove the need for a rotating dome slit, and simplify robotic observatory designs.  For a rough price, a 12.5 ft AstroHaven dome is listed at about \$50,000 USD, including a base ring, which could enclose telescopes less than about 70 cm in aperture, depending on the configuration.  

Operational costs: The financial commitment needed to maintain an observatory will vary depending on the site and the complexity of the system.  Here, we focus on site rentals (or equivalent), with the understanding that basic maintenance may require an extra 10\% or more.   

CTIO has an minimum site rental fee of \$30,000 USD per year associated with observatories on Cerro Tololo.  For a robotic telescope, the minimum is determined by the observatory footprint, and is set to 1250 sq.~ft or less. The USO is subject to this minimum fee, or about \$82 USD per day.   Assuming approximately 220 nights with 8 hr of usable imaging, the hourly cost of the telescope is about \$17 USD per hour to cover the site fee, which includes power and basic mountain maintenance. 

The SMARTS telescopes are operated by an open collaboration.  The use of the 0.9 metre costs \$600 USD per night, and users who have contributed over \$25,000 USD gain voting rights on the consortium.  The yearly rental costs and maintenance of the facility on Cerro Tololo is approximately \$80,000 USD (est.), justifying the nightly usage fee.  Similarly, the WIYN 0.9 metre is also open to international partners.  Use of this facility is approximately \$800 USD per night.  We note that the SMARTS and WIYN telescopes have different instrumentation, the latter equipped with a multi-chip, high-resolution half-degree imager. 

Occasionally telescopes at premier observing sites require new principal operators/consortia; the 1.3 metre on Cerro Tololo is an example.  Everything is already in place and functional for classic observing.  Making the facility robotic, if desired, would come with capital costs (estimated to be between \$100,000 to \$200,000 USD).  Scaling from the 0.9 metre, the yearly site rental is estimated to be approximately \$100,000 USD per year.  

Overall, to install a new, robotic 50 cm telescope on a premier site such as Cerro Tololo costs under \$200,000 USD (including shipping, computer equipment, and weather sensors). Operating costs would be between \$30,000 and \$35,000 USD per year, depending on the maintenance requests.  For much more significant capital costs, a turn-key metre-class telescope could be installed and remain subject to the minimum site fee (although this would require confirmation from the mountain director prior to installation). A different approach would be to take on ownership of an existing telescope, which could yield a much larger aperture size for a low capital cost but potentially much higher operational costs. 

How does ownership or partnerships compare with using a pay-for-service access such as iTelescope?  At the New Mexico Skies location, using a 50 cm telescope without a regular payment plan would cost approximately \$250 USD per hour of imaging if the Moon illumination is $<25\%$.  If the Moon is $>75\%$, the rate is reduced to \$124 USD per hour. 
If, instead, a payment plan of \$1000 USD is selected, billed every 28 days, the 50 cm telescope hourly costs are reduced to \$72 per hour for dark time and \$36 per hour for bright time. A payment plan of \$490 USD per 28 days increases the hourly rate by about 20\%.  Cheaper plans increase the rates further and approach the no-plan rate.  The website does note that an educational discount is available. 

Overall, a university can operate a telescope at a premier site for a cost that is comparable to or lower than using a company such as iTelescope, depending on the telescope's usage. Ownership comes with a number of additional advantages, such as direct control over the instruments and telescope scheduling.  However, it also comes with the need to handle the logistics of a remote facility and carries extra risk should a major problem occur.  

\begin{center}\bf Education and HQP Opportunities \end{center}

Small and moderate aperture telescopes offer a variety of educational opportunities, and can provide research experiences for high school students to undergraduates to postdocs, as well as public engagement.  They also create platforms for instrument development and hands-on experiences for HQP. Below is an incomplete list of examples using different approaches to include observations into curricula; these are only intended to be sketches of any particular option. We further realize that many universities are already employing some version of the following sketches or have a different approach entirely.  The hope is that this list can help to share ideas and to challenge us to think of ways to incorporate observing into a wide scope of classes, provided it is appropriate to do so.

Observational planning could be included in large undergraduate service courses.  A way to do this might start by presenting a list of potential targets to the class.  The students could then discuss in groups which targets they would most like to observe, as well as the types of information that might be learnt by observing the chosen objects.  The choice of observing targets could also be integrated into homework, projects, or labs.  If a large number of options are available, an in-class voting scheme, with iClicker or equivalent, would aid in selecting the final targets.  Teaching assistants could then acquire the images or spectra (which itself is a training experience for graduate students), followed by displaying them in class for further discussion. Lab sections could use the data for basic image analysis tasks, as well.  Such a model is adaptable to involve citizen science initiatives, International Asteroids Warning Network (IAWN) campaigns, exoplanet follow up, etc.
    
Observations from a premier site could be used in an upper-level undergraduate course on astronomical observations, which might involve observational planning, data acquisition, data reduction, and data analysis. The use of a telescope at a major observing site has the potential to facilitate many different types of observations and could ultimately contribute to ongoing research efforts.  The UBC telescope at CTIO is already incorporated into of UBC's astrophysical lab course ASTR 405.  

All incoming MSc students could be required to complete one project using a small to moderate aperture telescope.  This could be done to address a science question of their design or as part of one of the ongoing projects at the facility, including citizen science initiatives or other programs connected with undergraduate courses. 

 Departments and their universities could commit to sending each astronomy graduate student to the department's research observing site at least once. This might be incorporated into routine maintenance visits, educational programs, or part of an observing run.    While hands-on experience is possible using local telescopes, having a facility at a major research observatory complex provides students with experiences that are becoming harder to facilitate with dedicated telescope operators and queue observations.  As an example, Indiana University's astronomy department is committed to sending each graduate student to the WIYN 0.9 metre or another research facility. 

Undergraduate observing site visits could also be facilitated by a university, including smaller teaching institutions, with the opportunity to directly use the telescope. The WIYN 0.9 metre is again an example of an observatory that accommodates this while regularly conducting research.   

The Canadian Collaborative Occultation Network (CanCON) provides an example of how projects using small telescopes can engage high school students in astronomical research. To date, about 50 students and 7 teachers at five schools are participating in CanCON. Two journal papers have already resulted from CanCON data, and one of those papers has a high school teacher and several of his students as co-authors.

\begin{center}\bf Recommendations \end{center}

Small and moderate aperture telescopes can facilitate a diverse set of science programs, including exoplanets, binaries, variability, Solar System minor bodies, low-surface brightness structures, spectroscopy, microlensing, and large sky surveys, as examples. At premier sites, small telescopes can outperform larger observatories that are located in areas with poorer sky conditions. Such facilities can further be used for follow up science that is not as easily scheduled on larger instruments.  The costs of small to moderate facilities are relatively low, certainly in comparison with major observatories, but also in comparison with laboratories that can be found across university departments.  We recommend the following for advancing Canadian astronomy and astronomy education.

\begin{itemize}
    \item Coordinate with the NRC, the Tri-Council, and the Canadian Foundation for Innovation to build additional support for providing funding opportunities to establish domestic and international small and moderate aperture observatories, including both capital and operational costs (with site rental costs allowed).  While the UBC Southern Observatory is currently being upgraded thanks to an NSERC RTI, the placement of the observatory at CTIO created uncertainty in whether the facility itself is a Canadian asset (and therefore questioned its eligibility -- it was ultimately allowed). The costs involved also make direct funding through current programs difficult, as the observatories are typically too expensive to be funded in whole through existing programs but not necessarily expensive enough for major initiatives.  There is also not a clear funding program that would support the ongoing operational costs of such a facility. 
    \item CASCA should establish and maintain a small to moderate telescope expression of interest database.  This could include the scope of the facility, the desired range of science and/or education programming, and any details about an institute's desired partnership. This could help universities coordinate resources and establish programs and facilities with ongoing institutional support.  It would further help to involve small universities in any related initiative.
    \item Work with the NRC to make existing facilities more accessible to the astronomical community for research, including smaller institutions. This could involve, for example, automating the Plaskett and/or providing travel funds for supporting classical observing modes.  The latter could be of particular importance to HQP.  
    \item Work toward establishing an Arctic observatory. Small to moderate aperture telescopes could be highly impactful at a site location such as PEARL and would not require the same infrastructure and support as a large-aperture telescope. Close cooperation with NRC and the DND, who will have interests in orbital debris characterization, will be vital over the next decade to advance such a facility. 
    
\end{itemize}

%




\begin{lrptextbox}[How does the proposed initiative result in fundamental or transformational advances in our understanding of the Universe?]

Small to moderate aperture telescopes enable fundamental astronomy that, in some cases, cannot easily be conducted by large telescopes (either brightness constraints or scheduling conflicts).  They support a broad range of science cases in addition to HQP training, and can be used to support programs carried out at major observatories.

\end{lrptextbox}

\begin{lrptextbox}[What are the main scientific risks and how will they be mitigated?]

The main scientific risks are from throttling HQP training, support science, and innovative research programs by failing to support small and moderate facilities.  

\end{lrptextbox}

\begin{lrptextbox}[Is there the expectation of and capacity for Canadian scientific, technical or strategic leadership?] 

Canada can enter into MOU's with major observing site locations to establish new observatories or to assume operations of an existing telescope, requiring leadership within the Canadian community and international cooperation between Canadian universities and, for example, AURA.

\end{lrptextbox}

\begin{lrptextbox}[Is there support from, involvement from, and coordination within the relevant Canadian community and more broadly?] 

As highlighted in the WP, there are multiple small to moderate telescope programs being led by Canadian PIs.  

\end{lrptextbox}

\begin{lrptextbox}[Will this program position Canadian astronomy for future opportunities and returns in 2020-2030 or beyond 2030?] 

We argue that expanding support for small and moderate aperture telescopes will increase the opportunities for future science programs.  Supporting an Arctic small and moderate telescope facility would be of high value and would include international interest. 

\end{lrptextbox}

\begin{lrptextbox}[In what ways is the cost-benefit ratio, including existing investments and future operating costs, favourable?] 

The capital and operational costs of modern small and moderate aperture telescopes are low and can be supported by a single large university or a consortium of smaller institutes, although ongoing support from funding agencies is also needed.   A small to moderate telescope at a good observing site can outperform a large telescope at a moderate or poor quality site. 

\end{lrptextbox}

\begin{lrptextbox}[What are the main programmatic risks
and how will they be mitigated?] 

The main risks are through maintenance of the facilities and the potential for unplanned increases in site rental fees.  These can be mitigated through contingency funds set aside for an observatory and active communication with mountain directors. Observatories, whether domestic or international, should also ensure that there is ongoing and clear dialogue between the facility and the local community. 

\end{lrptextbox}

\begin{lrptextbox}[Does the proposed initiative offer specific tangible benefits to Canadians, including but not limited to interdisciplinary research, industry opportunities, HQP training,
EDI,
outreach or education?] 

Small and moderate aperture telescopes are well suited for engaging the public through outreach, secondary school STEM initiatives, undergraduate research opportunities, and HQP training.  Domestic observatories can foster deep connections with their local community, and international sites create important training opportunities for HQP.

\end{lrptextbox}

\bibliographystyle{unsrt}
\raggedright
\footnotesize
\bibliography{example}

\end{document}